\documentclass[a4paper,11pt]{article}
\usepackage{jheppub}

\usepackage{verbatim}
\usepackage{float}
\usepackage[small]{caption}
\usepackage{subfig}
\usepackage{booktabs}

\newcommand{\bra}[1]{\langle #1|}
\newcommand{\ket}[1]{|#1\rangle}

\def\Tr{{\rm{Tr}}}

\def\Rea{{\rm{Re}\,}}

\newcommand{\be}{\begin{equation}}
\newcommand{\ee}{\end{equation}}
\newcommand{\bea}{\begin{eqnarray}}
\newcommand{\eea}{\end{eqnarray}}
\newcommand{\bal}{\begin{align}}

\newcommand{\enl}{\end{align}}

\def\s{\sigma }

\def\ov{\over  }

\title{The one-loop and Sommerfeld electroweak corrections to the Wino dark matter annihilation}
\author{Andrzej Hryczuk,}
\author{Roberto Iengo}
\affiliation{SISSA and  INFN, Sezione di Trieste \\
 via Bonomea 265, 34136 Trieste, Italy}
 
 \emailAdd{hryczuk@sissa.it}
 \emailAdd{iengo@sissa.it}

\abstract{We compute the present-day Wino dark matter annihilation cross-section including the one-loop radiative corrections together with the fully treated electroweak Sommerfeld effect. We discuss what is the consistent way of incorporating these two corrections simultaneously and why simply using the running coupling constants values at the Wino mass scale is not correct. The results show that up to a few TeV scale the full one-loop computation makes the cross-section smaller up to about 30\% with respect to the Sommerfeld enhanced tree level result and are considerably larger than the tree or one-loop level without the Sommerfeld effect.}


\arxivnumber{1111.2916}

\begin{document}

\maketitle
\flushbottom

\section{Introduction}
The annihilation cross section is one of the key ingredients of computing the thermal relic density of the dark matter, as well as the indirect detection signals. Due to the fact, that until recently the observational data for the relic density and cosmic rays spectra were far from being accurate, most of the work in the literature uses only its tree level value. In recent years however, some work has been done in order to go beyond and include one-loop corrections \cite{Boudjema:2005hb,Baro:2007em,Baro:2009na}, as well as Bremsstrahlung processes \cite{Kachelriess:2007aj,Bell:2008ey,Dent:2008qy,Kachelriess:2009zy,Ciafaloni:2010qr,Ciafaloni:2010ti,Bell:2011eu,Ciafaloni:2011sa,Bell:2011if,Garny:2011cj,hep-ph/0507229,arXiv:0710.3169}. It has been found that including higher order processes can lead to rather large corrections and also may significantly alter the cosmic rays spectra.

Another effect studied recently in this context is the so-called Sommerfeld enhancement \cite{Sommerfeld}, that is the modification of the wave function of the incoming non-relativistic particles due to their mutual interaction (which is non-perturbatively treated). It has been shown to be a very important modification in a number of models with new dark forces (see e.g. \cite{Finkbeiner:2010sm} for a recent review), and even in more standard cases like in the Minimal Supersymmetric Standard Model (MSSM) \cite{Hisano:2002fk,Hisano:2003ec,Hisano:2004ds,Hisano:2006nn,Hryczuk:2010zi,Hryczuk:2011tq}. 

In this work we describe how to incorporate these corrections simultaneously and give the full one-loop and Sommerfeld corrected results for the case of the dark matter being a fermion living in the adjoint representation of $SU(2)_W$. This is the case of the pure Wino neutralino, but it is also interesting case per se (see e.g. the Minimal Dark Matter model \cite{Cirelli:2005uq}). It is also a good starting point for possible extensions because its relative simplicity makes more clear the description of the various effects. Another, more phenomenological reason to study this case is that it is precisely the one in which the Sommerfeld effect in the MSSM is the most important.

We start the discussion from presenting the model and reviewing the Sommerfeld effect. In section \ref{EWcorrections} we compute the radiative corrections and present the results. In section \ref{crossection} we compute the full cross-section (including the Sommerfeld enhancement). Finally we very briefly mention the possible phenomenological applications and give our conclusions in section \ref{conclusions}.

\subsection{The model}

We consider a model in which a Majorana fermion plays the role of the dark matter. We call it $\chi^0$ and assume that it belongs to the adjoint representation of the $SU(2)$ subgroup of the electroweak $SU(2)\times U(1)$. The other two members of the triplet can be combined to be described as a charged Dirac fermion and its anti-fermion which we call $\chi^\pm$. We assume that this triplet of fermions is massive due to an explicit mass term, that is present independently of the Higgs mechanism that might give mass to the weak vector bosons, and in fact we assume that these fermions do not interact with the Higgs field (if any). Within this setup, $\chi^0$ interacts only with the charged weak-interaction vector bosons $W^\pm$ and its charged partners only interact with the $W^\pm$, and with the $Z$ and $\gamma$.

We will be interested in the the mass $m$ of the Wino up to a few TeV. For the computation of the radiative corrections we also assume that the charged fermions of the multiplet have a mass which is higher of a negligible amount with respect to  the TeV scale, however we will include this difference in the Sommerfeld effect.\footnote{The mass difference comes from radiative corrections and is of the order of $\delta m=0.2$ GeV \cite{Cheng:1998hc}.}

In our model, at the tree level there is only one possible annihilation channel:\footnote{In the pure Wino scenario in the MSSM there are additional annihilation channels. However, in the case in which we are most interested in, i.e. $\chi^0$ having a mass in the TeV range, this channel is by the far dominant one.}
\be
\chi^0\chi^0 \to W^+ W^-.
\ee

However, at a higher order it is possible that the $\chi^0\chi^0$ pair becomes a (real or virtual) $\chi^+\chi^-$ pair which subsequently annihilates:
\be
\chi^+\chi^- \to W^+ W^- \qquad {\rm or}\qquad \chi^+\chi^- \to ZZ,Z\gamma,\gamma\gamma.
\ee
This sequential process is formally of higher order, but it can be enhanced by the two-channel version of the Sommerfeld effect (see section \ref{secSE}). In this case it can be effectively of the same as the tree level process and has to be included.

Moreover, since we consider the radiative corrections which provide an order $\mathcal{O}(g^4)$ correction to the amplitude and 
corresponding to an order $\mathcal{O}(g^6)$ term to the cross-section, we have also to include the annihilation in three final particles, i.e.
\be
\chi^0\chi^0 \to W^+ W^-Z,\ W^+ W^-\gamma
\ee
and, by Sommerfeld effect, also
\begin{equation}
\chi^0\chi^0 \to \chi^+\chi^- \to  W^+  W^- Z,\ W^+  W^- \gamma \, .
\end{equation}

We are interested in the case when annihilating particles are non-relativistic, therefore, we can take $\sigma_0$ to be the nominal cross-section for the annihilation at rest within a negligible relative error $\mathcal{O}(v^2)$, that is the relative variation of the Mandelstam variables $s$ and $t$ averaged over the angles. In this case the two incoming neutralinos, being Majorana fermions, form an $s$-wave spin-singlet. 
This is a very good approximation for dark matter particles in the halo today, and it allows for a great simplifications of the computations. Firstly, because in this case the initial pair (being Majorana fermions) have to be in a $s$-wave spin singlet. Secondly, the kinematics simplifies, since the annihilation becomes like a decay of a particle with the mass $2m$.

This approximation however sets limits on the usage of the results for the relic density calculations. Although at freeze-out the dark matter particles are still non-relativistic, their velocity is about $v\sim 0.3$. On the other hand, in order to get accurate results in this case one needs not only to generalize this computations, but also include the $p$-wave, which is beyond the scope of this work.\footnote{For some results including one-loop corrections to relic density computations, however without the fully treated non-perturbative Sommerfeld effect see refs. \cite{Baro:2007em,Baro:2009na}.}

\subsection{The Sommerfeld enhancement}
\label{secSE}

The Sommerfeld enhancement (for a recent review see \cite{Hryczuk:2010zi} and references therein) of a process 
is usually stated to give the full cross-section of the process in terms of
 the multiplication of two factors: 
\be
\label{secross}
\sigma=S(v) \sigma_0
\ee
where $S(v)$ is the velocity-dependent Sommerfeld factor, which represents the possible increase (decrease) of the flux of the incoming particles,
due to their mutual attraction (repulsion) and $\sigma_0$ is the nominal cross-section. Typically as $\sigma_0$ one uses the tree level value, but if one wants to incorporate loop corrections, it can be also computed at higher order in perturbation theory.\footnote{This is true as long as the annihilation process is short distance one, so that  the long distance Sommerfeld effect is decoupled and can be treated separately.} In our work we are going to compute $\sigma_0$ up the order $\mathcal{O}(g^6)$. Whereas we compute the nominal annihilation at rest, we still keep into account the velocity in the Sommerfeld enhancement, since in this process the momentum transfer can depend substantially on it.

The Sommerfeld enhancement factors are computed by a formalism using a system of  non-relativistic Schr\"{o}dinger equations. In the case at hand, there are two possible channels through which the annihilation can take place: $(\chi^0,\chi^0)$ and $(\chi^+,\chi^-)$. The $\chi^0$ pair can only become a (real or virtual) $\chi^\pm$ pair by exchange of $W^\pm$, whereas the  $\chi^\pm$ pair  can self-interact by exchanging $Z$ or photon and also come back to the $\chi^0$ pair by exchange of $W^\pm$.

This dynamics can be represented in terms of Feynman diagrams (see ref. \cite{Iengo:2009ni}). 
For small velocities, it is well described by the non-relativistic evaluation of ladder diagrams, where the steps of the ladder are 
either $W^\pm$, $Z$ or $\gamma$ and the lateral bars are $\chi^0$ or $\chi^\pm$.  
In the non-relativistic approximation (neglecting spin-orbit effects) the total spin of the two-body system is conserved, therefore both the pairs $\chi^0\chi^0$ and $\chi^+\chi^-$ are in a $s$-wave spin-singlet. Summing the ladder diagrams is equivalent to solving the coupled Schr\"{o}dinger equations for the two body (reduced) wave functions

Since in computing one-loop radiative corrections there are also diagrams containing the exchange of $W$ in the form of a single ladder, 
one has in this case to subtract the non-relativistic part which is already included in the solution of the Schr\"{o}dinger equations. For large 
$m$, this non-relativistic part can be quite large $\mathcal{O}({g^2\ov 4\pi} {m\ov m_W})$. The solution of the Schr\"{o}dinger equations represents a non-perturbative 
re-summation of those large terms and this is one reason for including the Sommerfeld effect.

 We call $\varphi^{0}(x)$ and $\varphi^\pm(x)$ the $s$-wave reduced wave functions for the $\chi^0\chi^0$ and $\chi^+\chi^-$ pair respectively
($x=r p$, $p=m v$, $v$ being the $\chi^0$ velocity).
In the low velocity approximation, the spin-singlet  $\chi^0\chi^0$ state is described by 
\be
\label{ini0}
\varphi^{0}(x) a^\dag_\uparrow a^\dag_\downarrow \ket{0} ,
\ee
and the spin-singlet $\chi^+\chi^-$ state by
\be
\label{iniC}
\varphi^\pm(x){a^\dag_\uparrow b^\dag_\downarrow-a^\dag_\downarrow b^\dag_\uparrow  \ov\sqrt{2}} \ket{0},
\ee
$a^\dag_{\uparrow,\downarrow} (b^\dag_{\uparrow,\downarrow})$ being particle (anti-particle) creation operators at rest, for a given spin-projection. 

Note the identities, for any Dirac matrix $M$:
\bea
\bra{0} \bar\chi^0 M \chi^0 a^\dag_\downarrow a^\dag_\uparrow \ket{0} &=& {1\ov  (2\pi)^3}\Tr\left[M{1+\gamma_0\ov 2}\gamma_5\right],
\\
 \bra{0} \bar\chi^+ M \chi^+{a^\dag_\downarrow b^\dag_\uparrow -a^\dag_\uparrow b^\dag_\downarrow \ov\sqrt{2}} \ket{0}  &=&{1\ov \sqrt{2} (2\pi)^3}\Tr\left[M{1+\gamma_0\ov 2}\gamma_5\right].
\eea
The coupled equations are ($\mathcal{E}=p^2/m$ is the energy of the incoming pair and $\delta m=m_{\pm}-m$):
\begin{eqnarray}
 \partial^2_x\varphi^{0}(x)\!\!\!&+&\!\!\!\varphi^{0}(x)+\frac{\alpha}{v}{e^{-{m_W\ov m} {x \ov v}} \ov x} \varphi^\pm(x)=0, \\ 
\partial^2_x\varphi^\pm(x)\!\!\!&+&\!\!\!q\left(1-\frac{2\delta m}{\mathcal{E}} \right)\varphi^\pm(x) \nonumber \\
\!\!\!&+&\!\!\!q\left[\left(c_w^2\frac{\alpha}{v}{e^{-{m_Z\ov m} {x \ov v}} \ov x}
+ s_w^2\frac{\alpha}{v}{1\ov x} \right)\varphi^\pm(x)
+\sqrt{2}\frac{\alpha}{v}{e^{-{m_W\ov m} {x \ov v}} \ov x} \varphi^{0}(x)\right]=0.
\end{eqnarray}
Here $q=\frac{m+\delta m}{m}$, $c_w=\cos\theta_w$ and $s_w=\sin\theta_w$, with $\theta_w$ being the Weinberg angle and $\alpha=g^2/(4\pi)$ where $g$ is the $SU(2)_L$ coupling. We take the values at the electroweak scale computed in the $\overline{MS}$ scheme \cite{PDG}: $s^2_w=0.23116$ and $g= 0.65169$.
One has to require the boundary conditions such that the $\varphi^0$ is incoming and outgoing, whereas $\varphi^\pm$ is only outgoing (if $2\delta m\leq \mathcal{E}$) or exponentially decaying (if $2\delta m > \mathcal{E}$).

As we explain below, in our case velocity is too small to allow for on-shell $\varphi^\pm$, hence the relevant case is only the latter one and the boundary conditions are
\be 
\lim_{x\to\infty}\left(i \varphi^{0}(x)- \partial_x \varphi^{0}(x)\right)=-e^{-i x}, \qquad \lim_{x\to\infty} \varphi^{\pm}(x)=0.
\ee
With this normalization, we call the \textit{amplitude} Sommerfeld factors:
\be
\label{seampfact}
s_0\equiv  \partial_x\varphi^0(x)|_{x=0} , \qquad s_\pm\equiv  \partial_x\varphi^\pm(x)|_{x=0} ,
\ee
then the (Sommerfeld enhanced) amplitudes of the annihilation processes for any Standard Model (SM) final state are:
\be
\label{sdef}
A_{\chi^0\chi^0\to {\rm SM}} = s_0 A^0_{\chi^0\chi^0\to {\rm SM}}+s_\pm A^0 _{\chi^+\chi^-\to {\rm SM}}.
\ee

In the literature one often finds the one channel version of the Sommerfeld. In this case one would have a single wave function $\varphi(x)$  and  the Sommerfeld enhanced the cross-section would be given by  eq. (\ref{secross}) 
with
\begin{equation}
S(v)=|s|^2, \qquad s=\partial_x\varphi(x)|_{x=0} .
\end{equation}
However, if there are more channels, one has to use the formula for the enhanced amplitude  eq. (\ref{sdef}). Note that by taking the modulus square of eq. (\ref{sdef}) there is also a cross term, which was neglected in the previous works on the Sommerfeld effect.

The result for the Sommerfeld factors depends strongly on the mass-splitting between $\chi^0$ and $\chi^\pm$; in fact, also the computation is somewhat different if the total energy of the $\chi^0$ pair is greater or smaller than twice the $\chi^\pm$ mass. However, taking the mass splitting to be of the order of $0.2$ GeV, and the velocity of the order of present day dark matter velocity $v\sim 10^{-3}$, even for $m$ being a few TeV,  the production of real $\chi^\pm$  from the $\chi^0$ pair is not allowed. In this case (i.e. far below the $\chi^+\chi^-$ threshold) $s_{0,\pm}$ nearly does not depend on the velocity \cite{Hryczuk:2010zi,Slatyer:2009vg}.

It is worth noting, that for small velocities the Sommerfeld effect at one loop level (i.e. not summed over all orders), gives:
\begin{equation}
\label{eq:oneloopS}
s_0=1, \qquad s_\pm= \sqrt{2} {g^2\ov 4 \pi}{ m\ov m_W}.
\end{equation}
This result is recovered from the full solution of the Schr\"{o}dinger equations when one takes the limit of small $v{ m\ov m_W}$ and small ${g^2\ov 4 \pi}{ m\ov m_W}$. However, since for large enough $m$ this does not hold, the full numerical computation is needed.

\section{Electroweak corrections}
\label{EWcorrections}

In this section we will present the computation and the results for the radiative corrections to the annihilation amplitudes. 
In order to compute the cross section up to the order $\mathcal{O}(g^6)$ we need to include both the one-loop corrections and the real production. In particular, the Bremsstrahlung of a soft/collinear photons is crucial for the cancellation of the IR divergences, as we will discuss in detail in section \ref{IRdiv}.

We will organize the results in the following way. Firstly we discuss the issue of at what scale should we take the couplings. Then we consider the radiative corrections to the amplitudes 
$A_{\chi^0\chi^0\to W^+W^-}$, $A_{\chi^+\chi^-\to W^+W^-}$, $A_{\chi^+\chi^-\to ZZ,Z\gamma,\gamma\gamma}$. 

\subsection{The scale of the coupling}
\label{couplscale}

It is very important to understand at what energy scale the $SU(2)$ coupling $g$ and of the Weinberg angle $\theta_w$ should be taken. We argue, that both in the radiative corrections and the Sommerfeld effect computations we should take them at the electroweak (EW) scale.

In the computation of  the Sommerfeld factors $s_{0,\pm}$ one could wonder whether one should take $g$ at the scale of the neutralino mass $m$, since this sets the energy scale of the process. However, what matters in the Sommerfeld enhancement computation is in fact the scale of the momentum transfer between the incoming particles and this can be at most of the order of the vector boson mass.  

The detailed computation, both analytical and numerical, shows \cite{Iengo_unpublished} that the radiative corrections to the vertices $\chi\chi W$, with $\chi$ on-shell as appropriate in the non-relativistic case, at zero momentum transfer exactly compensate the effect of the $\chi$ wave-function renormalization. Therefore, there is no dependence on $m$ 
once taking the renormalization scale to be $m_W$, and this compensation persists quite effectively up to momentum transfer $\mathcal{O}(m_W)$. Therefore, since what remains to be considered is the $W$ wave-function renormalization, which is already included in the definition of the coupling
at the scale $m_W$, there are no appreciable corrections at all, if we take $g$ at the scale $m_W$. 

In fact, note that the use of the running coupling constant is appropriate for the processes which depend significantly on a \textit{single} large scale. For instance it would be appropriate to take the coupling at the scale $m$ for processes in which the momentum transfer to $\chi$ is also of the order of $m$.

In the case of the radiative corrections to the \textit{annihilation amplitude}, there is not a precise compensation of the radiative correction of the vertices $\chi\chi W$ with the $\chi$ wave-function renormalization because the internal $\chi$ lines are off-shell, and therefore we take into account these loops that give a further radiative correction not included into taking $g$ at the scale $m_W$. 

Note that, by computing the Feynman diagrams giving the vertex corrections and the wave function renormalization and fixing the renormalization at the EW scale, we are evaluating perturbatively how the coupling "runs" from its EW value. 

This is like expanding at the one-loop order the formula for the running coupling constant, except that we do not have  to include the $W$ wave-function renormalization, because 
it only depends on the square $W$-four-momentum which is equal to $m_W^2$ and therefore it is already inside the definition of the coupling at the scale $m_W$. 

Let us also recall that the standard use of the renormalization group techniques holds  in the "deep euclidean region" in which the external lines are quite off-shell.
In our case instead, the external particles are on-shell and therefore there occur not only the large log's related to the UV divergences but also large log's due to IR effects.
As we will discuss in the following, we do not attempt a re-summation of the large log's of various origin, and this is another reason why we do not attempt to use a kind of non-perturbative formula for the running coupling,
suitably modified to take off the $W$-wave function renormalization,
which would correspond to some partial re-summation of one subset only.

\subsection{The radiative corrections to $\chi^0\chi^0\to W^+W^-$}

We start the discussion from the one-loop corrections to $\chi^0\chi^0$ annihilation. Firstly we will discuss the method of doing the computations and the in section \ref{neutralinoresults} we will give the results.  The way we present them is in terms of the correction to the tree level amplitude: 
\begin{equation}
A=A_{{\rm tree}}\left(1+\frac{g^2}{(4\pi)^2} C_i(m) \right),
\end{equation}
where $C_i(m)$ are the coefficients corresponding to the diagram $i$. 

\subsubsection{The UV divergent diagrams}

\begin{figure}[h]
\centering
	\includegraphics[scale=0.7]{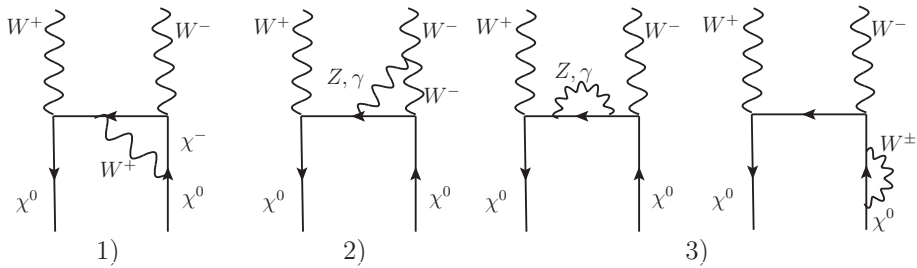} 
	\caption{The UV divergent diagrams for $\chi^0\chi^0\to W^+W^-$ process. The vertex corrections (diagrams $1$ and $2$) and the fermion wave-function renormalization (both diagrams are included in $3$). }
	\label{diag1}
\end{figure}

The UV divergent one-loop diagrams come from the vertex corrections and the fermion wave-function renormalization, as presented on figure \ref{diag1}.

For all these diagrams, we have done the computations using full analytical expressions with Feynman parameters and integrated analytically (using \textit{Mathematica}) over the first one and numerically over the second one.\footnote{In the approximation of annihilation at rest all the diagrams can be expressed as a linear combination of integrals with only two Feynman parameters, because in this case there are only two independent external momenta.} We took the Feynman-'t\,Hooft gauge for the $W$ propagator, which simplifies the computations, noting that $\chi$ is not coupled to the Higgs bosons and that there is no vertex with two $W$'s and one neutral unphysical Higgs. We used dimensional regularization and dropped the terms $\mathcal{O}(1/\epsilon)$ because they are taken into account in the renormalization at the scale $m_W$. In fact, the loop corrections to the coupling $g$ evaluated at the $m_W$ scale do contain the same $\mathcal{O}(1/\epsilon)$ terms, which therefore are part of the definition of the coupling at that scale. We also didn't include the $W$ wave-function renormalization of the final $W$'s for the same reason.\footnote{Except that we have to include the IR divergence of the $W$ wave-function renormalization due to the photon exchange, which is cancelled by a real photon emission, see section \ref{IRdiv}.}

\subsubsection{The UV finite diagrams and the IR divergence}

Besides the loops giving the radiative correction of the vertices and the $\chi$ wave-function renormalization, there are two other loops, 
which are not UV divergent.\footnote{The propagators and vertices of these diagrams give three powers of momentum in the numerator and eight powers in the denominator therefore the integration in four dimensions is convergent by power counting. In the case of diagram 5 the analytic integration on one parameter has been done using the PrincipalValue prescription, in order to discard the absorptive part, due to intermediate $W$'s being possibly on-shell. This part does not interfere with the tree diagram and thus would give a higher order contribution. 

The diagram containing the four vector boson vertex gives a vanishing contribution for the Wino annihilation at rest in a spin-singlet state. }

\begin{figure}[ht]
\centering
	\includegraphics[scale=0.7]{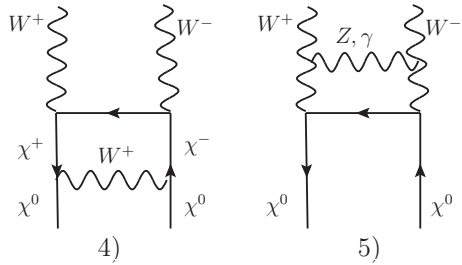} 
	\caption{The UV finite diagrams for $\chi^0\chi^0\to W^+W^-$ process.}
	\label{diag2}
\end{figure}

Diagram 4 represents a process in which the incoming $\chi^0$ pair goes to a virtual $\chi^\pm$ pair (which then annihilates in $W^\pm$) by  $W^\pm$ exchange (see figure \ref{diag2}, diagram 4). The contribution of this loop is very large when $m/m_W$ is large. In fact, it is recognized that it contains the first order  contribution
to $s_\pm$.\footnote{Divided by the $\sqrt{2}$, due to the difference in the normalization of the initial two-body state for neutralino and chargino pairs, see Eqns. \eqref{ini0} and \eqref{iniC}.}   This is seen because, as shown in ref. \cite{Iengo:2009ni}, the Sommerfeld effect comes by summing the non-relativistic part of the ladder  diagrams, and this diagram is precisely the first of the series.

More in detail, the statement for the Sommerfeld enhanced amplitude $A=s A_0$ comes from taking the amplitude as
the non-relativistic approximation of a sum of ladder diagrams  having $W$-propagators as steps. Schematically:
\be
A = \sum_{n=0}^\infty A_0\times (W_{{\rm step}})^n,
\ee
where $\times$ means a non-relativistic convolution in momentum space and $(W_{{\rm step}})^n \equiv W_{{\rm step}}\times\cdots\times W_{{\rm step}}$.  

Since the radiative corrections of $(W_{{\rm step}})^n$ have been already treated, one has to compute the radiative corrections of $A_{{\rm tree}}$,
that is replace
\be
A_{{\rm tree}}\to A_{{\rm tree}}+\sum_i D_i,
\ee
where $D_i$ are the various diagrams correcting $A_{{\rm tree}}$, 
but of course one must not include the contribution of non-relativistic approximation of the ladder diagrams 
that has been already taken into account. The contribution of the diagram 4 is then:
\be
D_4= A_{{\rm tree}}\times W_{{\rm step}} + A_{{\rm tree}} {g^2\ov (4 \pi)^2} C_4.
\ee
At a one-loop level $A_{{\rm tree}}\times W_{{\rm step}}=A_{{\rm tree}} {g^2\ov 4 \pi}{ m\ov m_W}$, and therefore
\be
A_{{\rm tree}} {g^2\ov (4 \pi)^2} C_4\equiv D_4 - A_{{\rm tree}} {g^2\ov 4 \pi}{ m\ov m_W}
\ee
has to be considered as the genuine contribution to the radiative correction to $A_{{\rm tree}}$. 

Diagram 5 is due to the exchange of $Z$ or $\gamma$ between the final $W^\pm$ (see figure \ref{diag2}). This loop is IR divergent in the part in which there is the photon exchange. This is the only IR divergence in the radiative corrections of $\sigma_0$, because the initial $\chi^0$ does not couple to the photon and there is no photon contribution to the  $\chi^0$ wave-function renormalization, moreover in the other loops at least one of the $\chi$ line is off-shell thus avoiding IR divergences. 

 \subsubsection{The total result for the radiative corrections due to the loops}
 \label{neutralinoresults}
  
\begin{figure}[t]
\centering
	\includegraphics[scale=0.45]{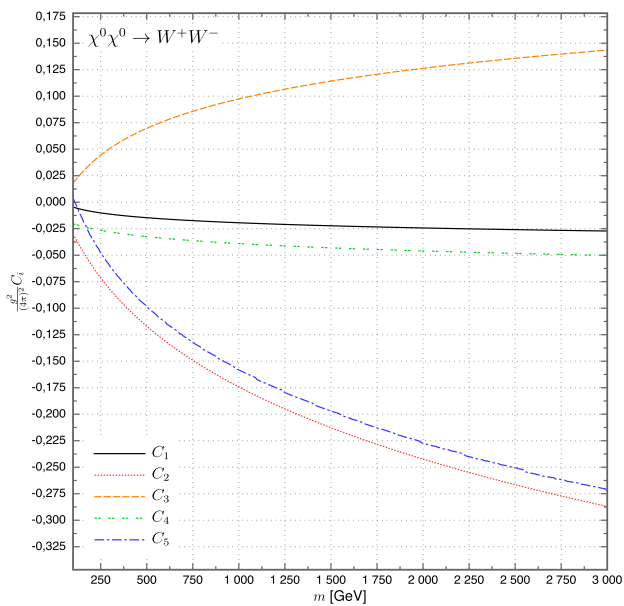} 
	\caption{The results for the one-loop correction to the amplitude of the $\chi^0\chi^0 \to W^+W^-$ annihilation. The total correction is obtained by summing all those contributions and including the real production. The $C_5$ contribution is made finite due to adding a small mass to the photon $m_\gamma=0.1$ GeV. In these result all the multiplicities of the diagrams were taken into account.}
    \label{res_diags}
\end{figure}

On figure \ref{res_diags} we present the results for the one-loop corrections coming from all the diagrams separately, in function of the DM mass being in range of 100 GeV - 3 TeV. One can see that the largest contributions come from diagrams $2$,$3$ and $5$, all of which contain photon exchange. Although the $C_5$ is IR divergent, we get a finite result by giving (in all the numerical calculations) a small (with respect to the TeV scale) mass $m_\gamma=0.1$ GeV to the photon. In reality it is of course massless, and indeed as we shall see the dependence on $m_\gamma$ will drop out in the final result. We will come back to this point in section \ref{IRdiv}, where we discuss the cancellation of the IR divergence by the inclusion of a real production.
 
\begin{figure}[t]
\centering
	\includegraphics[scale=0.7]{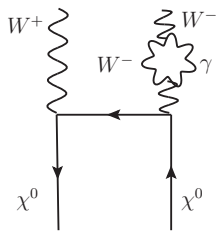} 
	\caption{The IR divergent diagram present in the gauge boson wave-function renormalization.}
	\label{diag3}
\end{figure}

 Actually, in order for the cancellation to be exact, we have also to take into account the IR divergent part of the virtual photon contribution to the W wave-function renormalization, 
 that is not included in the renormalization of $g$  at the scale $m_W$ (see figure \ref{diag3}). It gives a further contribution to $C$ which is: $s_w^2\, 4\log\bigl(\frac{m}{m_\gamma}\bigr)$. Including it in one-loop corrections gives finally
\be
C_{1-loop} =\sum_{i=1}^5C_{i}+ s_w^2 4 \log\left(\frac{m}{m_\gamma}\right).
\ee
 
 From the computation we get that $C_{1-loop}<0$ and also that it diverges for $m_\gamma\to 0$. The divergence is due to the graph $5$ described above and to IR divergent part of the W wave-function renormalization.

\subsubsection{The radiative correction due to the real production and the cancellation of the IR divergences}
 \label{IRdiv}

\begin{figure}[ht]
\centering
	\includegraphics[scale=0.7]{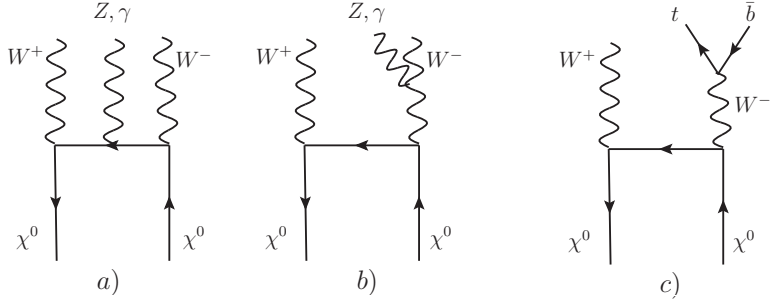} 
	\caption{The diagrams for real production of the gauge bosons and the production of the $t\bar b$ quark pair.}
	\label{diagrealt}
\end{figure}

As we already mentioned, doing the computation of $\sigma_0$ at the order $\mathcal{O}(g^6)$, we have also to add to $\sigma_0$ the production cross-section of $W^+W^-Z$ and $W^+W^-\gamma$,
which are of the same order (see figure \ref{diagrealt}). This has to be done on the level of the cross section. Hence, we start from a short review of the cross-section computation. If we call the amplitude $\mathcal{M}$, then the formula for the differential cross-section in our case reads:
\begin{equation}
\label{eq:dsvgen}
d\sigma=\frac{1}{4m^2 v_r} \left(\prod_i \frac{d^3k_i}{(2\pi)^3 2\omega_i}\right) \sum_{{\rm pol}}|\mathcal{M}|^2 (2\pi)^4 \delta^4(P-\sum_i k_i),
\end{equation}
where $v_r=2v$ is the relative velocity, $P=(2m,0,0,0)$ and $m$ is the mass of annihilating DM particle. The sum over polarizations gives:
\begin{equation}
\sum_{{\rm pol}} \epsilon_\mu \epsilon_\nu^*=-g_{\mu\nu}+\frac{k_\mu k_\nu}{m_{W,Z}^2}.
\end{equation}
for massive gauge bosons and
\begin{equation}
\sum_{{\rm pol}} \epsilon_i \epsilon_j^* = \delta_{ij}-\frac{k_i k_j}{\vec{k}^2},
\end{equation}
for the photon.

In the annihilation into two particles with the same mass $m_g$ the integration over the phase space gives:
\begin{equation}
\label{xsec2body}
\sigma_2 v=\frac{1}{64\pi}\sqrt{1-\frac{m_g^2}{m^2}}\sum_{{\rm pol}}|\mathcal{M}|^2.
\end{equation}

For the annihilation into three body final state, in the limit in which initial particles are in rest, the cross-section can be computed in a convenient parametrization with the use of Dalitz variables\footnote{In actual numerical computations we follow a more direct approach by integrating over the final energies, which we check to be equivalent; this is numerically more convenient but the formulae are too long and we don't write them here.}, $m_{ij}^2=(k_i+k_j)^2$:
\begin{equation}
d\sigma_3=\frac{1}{(2\pi)^3}\frac{1}{16(2m)^4}\frac{1}{v_r}\sum_{{\rm pol}}|\mathcal{ M}|^2dm_{12}^2 dm_{23}^2.
\end{equation}
The integration limits on these variables depend only on the masses and can be conveniently presented as \cite{Amsler:2008zzb}:
\begin{equation}
4m_1^2\leq m_{12}^2 \leq (2m-m_{3})^2 \qquad (m_{23}^2)_{min} \leq m_{23}^2 \leq (m_{23}^2)_{max},
\end{equation}
with
\begin{eqnarray}
&& (m_{23}^2)_{min}=(E_2+E_3)^2-\left(\sqrt{E_2^2-m_1^2}+\sqrt{E_3^2-m_{3}^2}\right)^2, \\
&& (m_{23}^2)_{max}=(E_2+E_3)^2-\left(\sqrt{E_2^2-m_1^2}-\sqrt{E_3^2-m_{3}^2}\right)^2.
\end{eqnarray}
Here $E_2=m_{12}/2$ and $E_3=(4m^2-m_{3}^2-m_{12}^2)/2m_{12}$ are the energies of particles 2 and 3 in the $m_{12}$ rest-frame.

In order to add these contributions to the one-loop corrections we define the coefficients $C_Z^{rp}$ and $C_\gamma^{rp}$ as:
\be
2{g^2\ov (4 \pi)^2}  c_w^2 \, C^{rp}_Z\equiv {\sigma_{W^+W^-Z}\ov \sigma_2^{{\rm tree}}}, \qquad 2  {g^2\ov (4 \pi)^2}  \s_w^2\,  C^{rp}_\gamma\equiv {\sigma_{W^+W^-\gamma}\ov \sigma_2^{{\rm tree}}}.
\ee
Note the factor 2 in the definitions, which makes these coefficients to be the \textit{corrections to the amplitude} coming from the real production. Using this we can write that the production cross-section provides a further correction:
\be
C_{1-loop+rp} =C_{1-loop}+ c_w^2 C_Z^{rp} + s_w^2 C_\gamma^{rp}.
\ee

The full $C_{1-loop+rp}$ coefficient should go to a finite constant for $m_\gamma\to 0$, which we find it is indeed the case. On the right plot of figure \ref{plot_00} we show the separate contributions from one-loop corrections and the real production to show that their sum is independent of $m_\gamma$.

\paragraph{Three body production involving $t$ quark.}

There is a further, though very small, contribution to the production cross-section at the order $\mathcal{O}(g^6)$: the processes involving
the $t$ quark in the final state
\bea
&&\chi^0\chi^0 \to W^-  t\bar d,\qquad \chi^0\chi^0 \to W^- t\bar s,\qquad \chi^0\chi^0 \to W^-  t\bar b, \nonumber \\
&&\chi^0\chi^0 \to W^+ \bar  td,\qquad \chi^0\chi^0 \to W^+  \bar ts,\qquad \chi^0\chi^0 \to W^+ \bar t b \nonumber
\eea
These processes are due to the couplings $W^+\to t\bar d,t\bar s,t\bar b$ and their conjugate.

Notice that the other processes where in the final state there are a charged $W$ and either a charged $l_i\bar l_j$ or a lighter charged $q_i\bar q_j$ pair 
must not be included, because they sum up to the total width of the charged $W$, and therefore are implicitly taken into account by unitarity
when one takes the approximation of considering $W$ as a stable particle. But the $top$ is more massive than the $W$ and therefore the charged $q\bar q$ pairs where one of the quarks is $t$ are not included in 
in the total width of the $W$ and have to be added to the correction.

Since the square of the coupling $W^+\to t\bar d$ is negligible with respect to the sum of the square of $W^+\to t\bar b$ and $W^+\to t\bar s$
(which all together add up to $g^2$),
and the masses of $b,s$ are negligible at our energy scale, by defining as before
\be
 2{g^2\ov (4 \pi)^2}  C_t \equiv {\sigma_{W^-t\bar b}+\sigma_{W^+\bar t b}\ov \sigma_2^{{\rm tree}}},
 \ee
 we get final result for the total correction to the tree amplitude
\be
C_{1-loop+rp+t} =\sum_{i=1}^5C_{i}+ s_w^2\left( 4 \log\left(\frac{m}{m_\gamma}\right)+C_\gamma^{rp}\right)+c_w^2 C_Z^{rp} +C_t\, .
\ee

In the numerical results, as we will see, the relative contribution of $C_t$ is very small, due to the fact that it does not contain any large Log's, which are present in the case of the production of three gauge bosons.

Note, that since the unphysical neutral Higgs is not coupled to $W^\pm$ we can use the Feynman-'t\,Hooft gauge for the vector bosons 
forgetting the unphysical Higgs. As for the physical Higgs, its coupling to $W^\pm$ is proportional to 
$g m_W$ and and therefore the (virtual or real) process involving it will be suppressed by a factor $m_W^2/m^2$ and
we neglect them.

\subsubsection{The total result for the annihilation of $\chi^0\chi^0$}
\label{chi0total}

\begin{figure}[t]
\centering
	\includegraphics[scale=0.4]{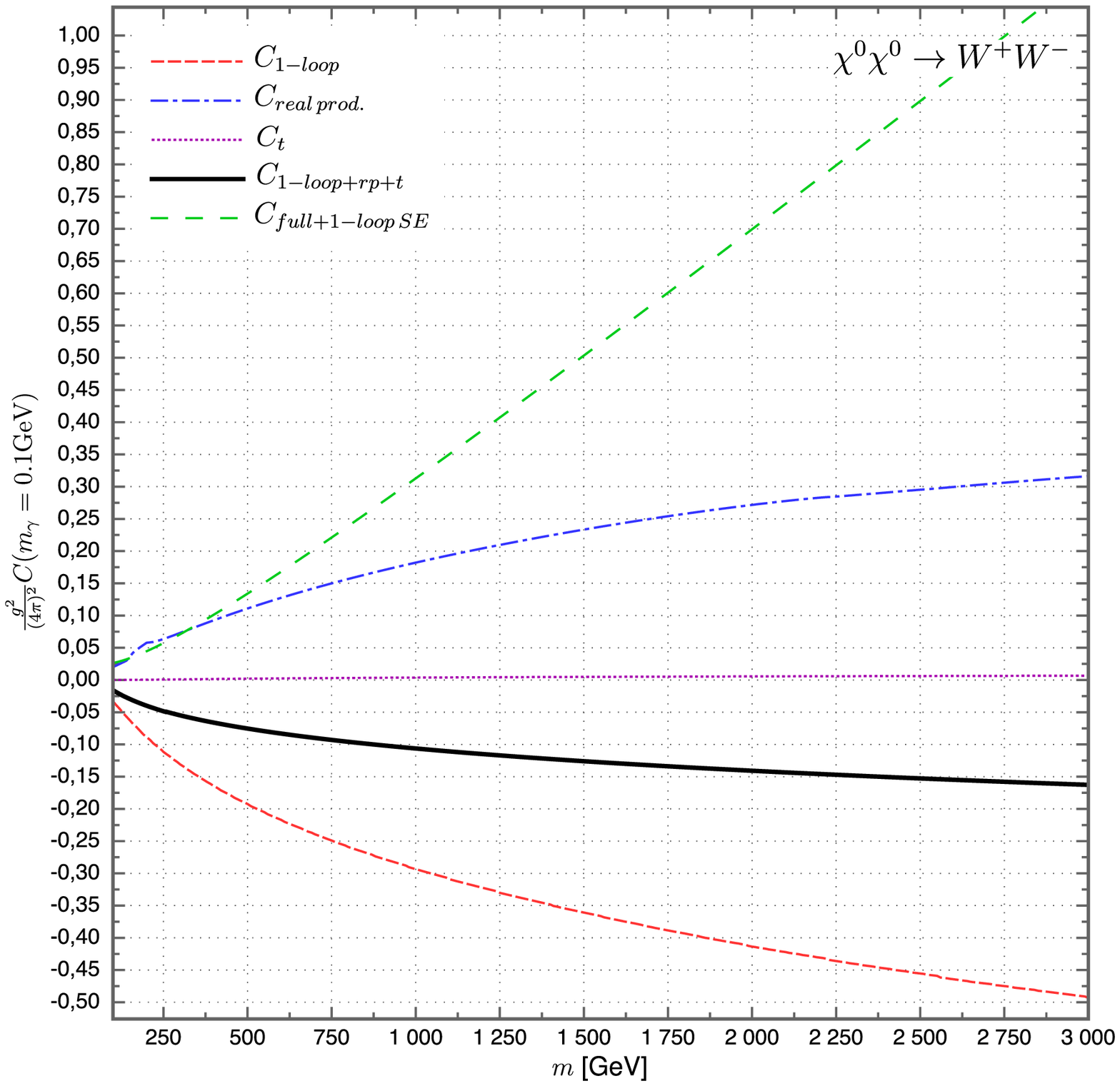} 
		\includegraphics[scale=0.4]{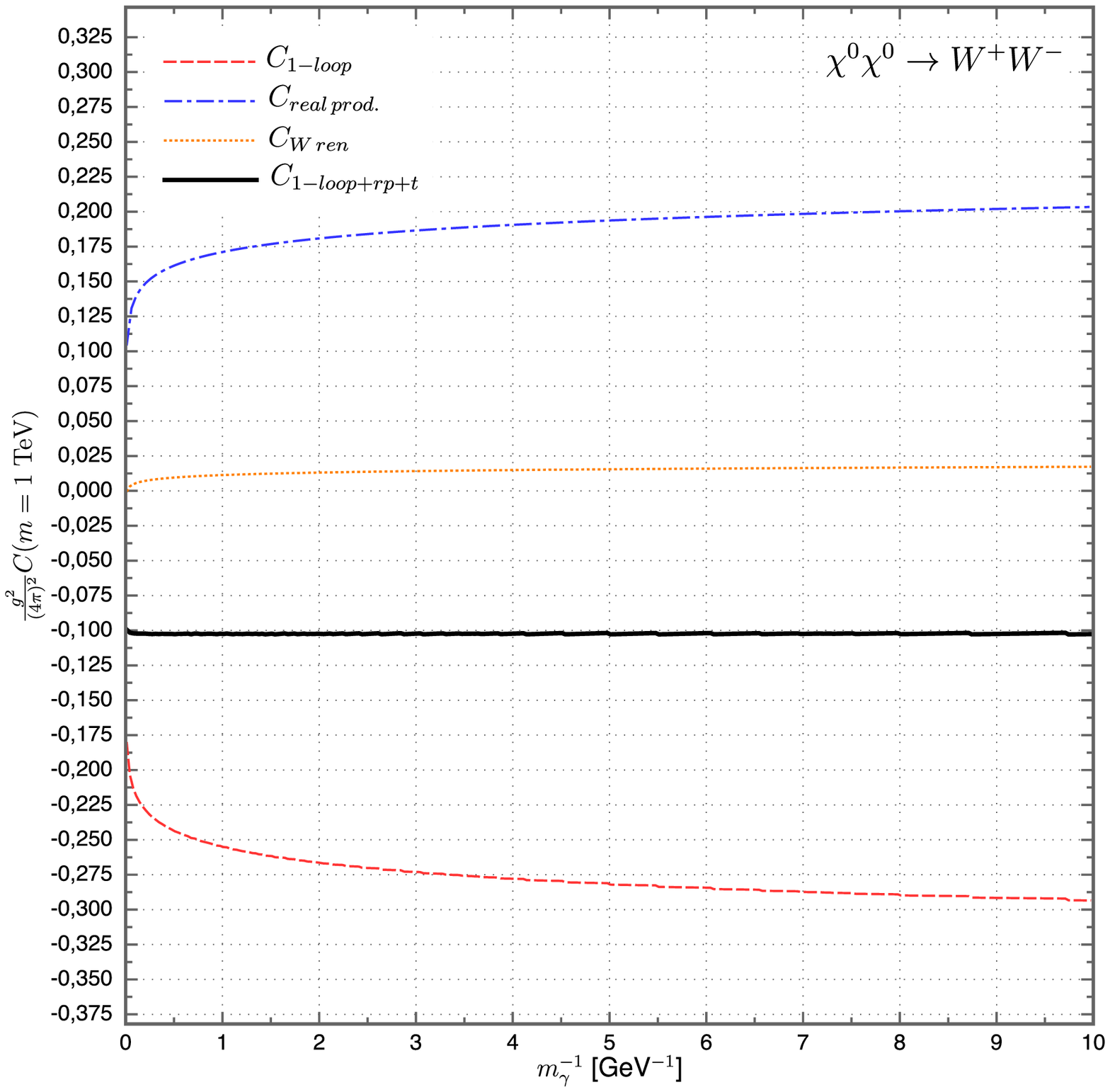} 
	\caption{\textit{Left plot:} the correction to the tree level $\chi^0\chi^0\to W^+W^-$ amplitude coming from loop corrections (dashed red line), real production (chain blue) and the $t$ quark (dotted violet). The full result is given by the solid black and sparse green lines (without and with the one-loop Sommerfeld correction, respectively). \textit{Right plot:} the dependence of the full result on the photon mass, for fixed $m=1$ TeV. A complete cancellation of the IR divergent terms can be seen, and that the full result is independent of $m_\gamma$.}
    \label{plot_00}
\end{figure}

We show the results for the full radiative corrections to the $\chi^0\chi^0$ annihilation \textit{amplitude} on the left panel of figure \ref{plot_00} (by now without the full Sommerfeld effect). One can see that, subtracting the one-loop Sommerfeld effect
(that will be non-perturbatively treated),
the total corrections (the solid black line) are significant, reaching over 15\% for the $m=3$ TeV, but still in the perturbative regime. 

We  also see that at a TeV scale the one-loop perturbative evaluation of the Sommerfeld effect is quite large and this is one of the reasons why the full non-perturbative treatment of this effect is needed.

In writing $\sigma=S\times\sigma_0$ we are assuming that, while $S$ is the non-perturbatively evaluated Sommerfeld effect, we can compute $\sigma_0$ 
by the standard Feynman diagrams of perturbation theory. However,  for growing $m/m_W$ the contributions to $\sigma_0$ of the diagrams $\mathcal{O}(g^6)$ becomes larger and larger, 
and the perturbative evaluation of $\sigma_0$ looses its meaning.
Indeed, the perturbative evaluation of the correction to $\sigma_0$ looks like to be border-line-reliable up to values of $m$ of a few TeV.

This fact is not surprising: when $m$ and therefore the overall scale of the process gets large as compared to $m_W$, 
the vector bosons resemble more and more to massless would-be gluons of an unbroken $SU(2)$, like an  $SU(2)$ version of QCD.
There occur large Log's of the ratio $m/m_W$, and powers of them, which are not related to the UV divergences 
(and therefore cannot be included in a standard renormalization group treatment). Therefore, for higher values of $m$, one would need to borrow from QCD sophisticate techniques 
of re-summation  of powers of large Log's or semi-empirical formulae. All that is beyond the scope of this work.

\subsection{The radiative corrections to $\chi^+\chi^-$ annihilation}

Due to the Sommerfeld effect the $\chi^+\chi^-$ annihilation gives a non-negligible contribution to the $\chi^0\chi^0$ annihilation process, which in fact can be of the same order as the direct process. Therefore, it is also important to compute the radiative correction to annihilation with $\chi^+\chi^-$ in the initial state. Because the computations are very similar to the $\chi^0\chi^0$ case, we don't discuss all the computations in detail, but rather stress the differences and present the final results. 

In this case, in the Feynman-'t\,Hooft gauge it occurs also the vertex of the charged unphysical Higgs with the vector bosons. However, in the same way as for the physical Higgs, its coupling is proportional to $gm_W$ and therefore the process involving it will be suppressed by a factor $m_W^2/m^2$ and we neglect it. 

\subsubsection{One-loop corrections to $\chi^+\chi^-\to W^+W^-$}

\begin{figure}[t]
\centering
	\includegraphics[scale=0.7]{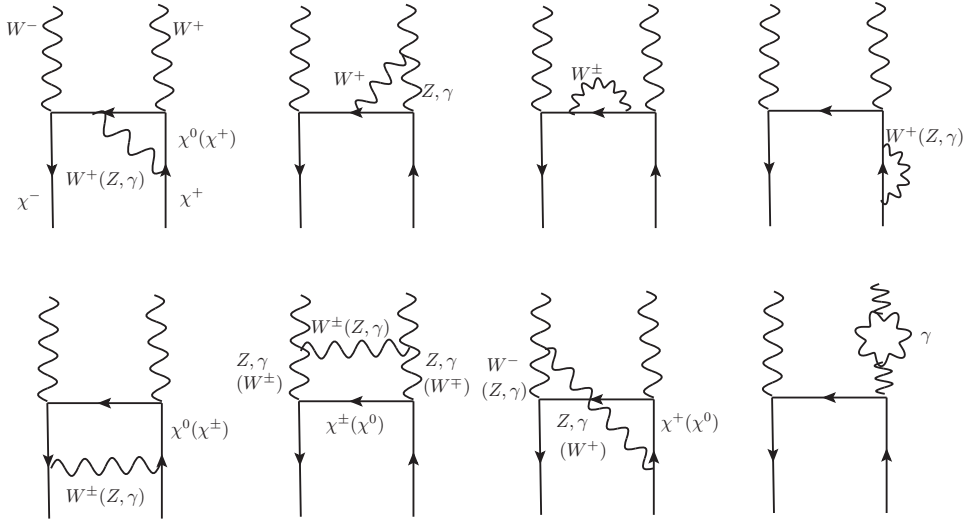} 
	\caption{The diagrams for the one-loop corrections to $\chi^+\chi^-\to W^+W^-$ annihilation.}
	\label{diagC}
\end{figure}

In the case of the annihilation of $\chi^+\chi^-$, since they are charged, there are more diagrams to be computed, see figure \ref{diagC}, the technique is however exactly the same.

Note also that due to the difference in the normalizations of the initial states (Eqns. \eqref{ini0} and \eqref{iniC}),  at the tree level $A^{{\rm tree}}_{\chi^+\chi^-\to W^+W^-}={1\ov \sqrt{2}}A^{{\rm tree}}_{\chi^0\chi^0\to W^+W^-}$.

\subsubsection{The radiative correction due to the real production}

\begin{figure}[t]
\centering
	\includegraphics[scale=0.7]{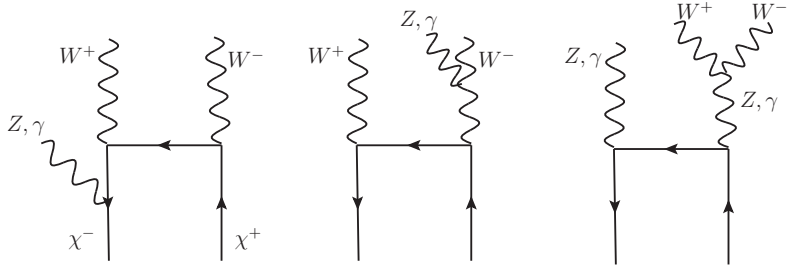} 
	\caption{The diagrams  the correction to the process $\chi^+\chi^-\to W^+W^-$ coming from  the real production of $Z,\gamma$.}
	\label{diagCwpwmw3}
\end{figure}

Also in this case the computations goes in the same way, except that now the initial state particles are coupled to $Z$ and $\gamma$, which gives the initial state Bremsstrahlung process (instead of internal one as in the $\chi^0\chi^0$ case). The diagrams to be computed are those on figure \ref{diagCwpwmw3}.

\subsubsection{The total result for the annihilation of $\chi^+\chi^-\to W^+W^-$}

\begin{figure}[t]
\centering
	\includegraphics[scale=0.4]{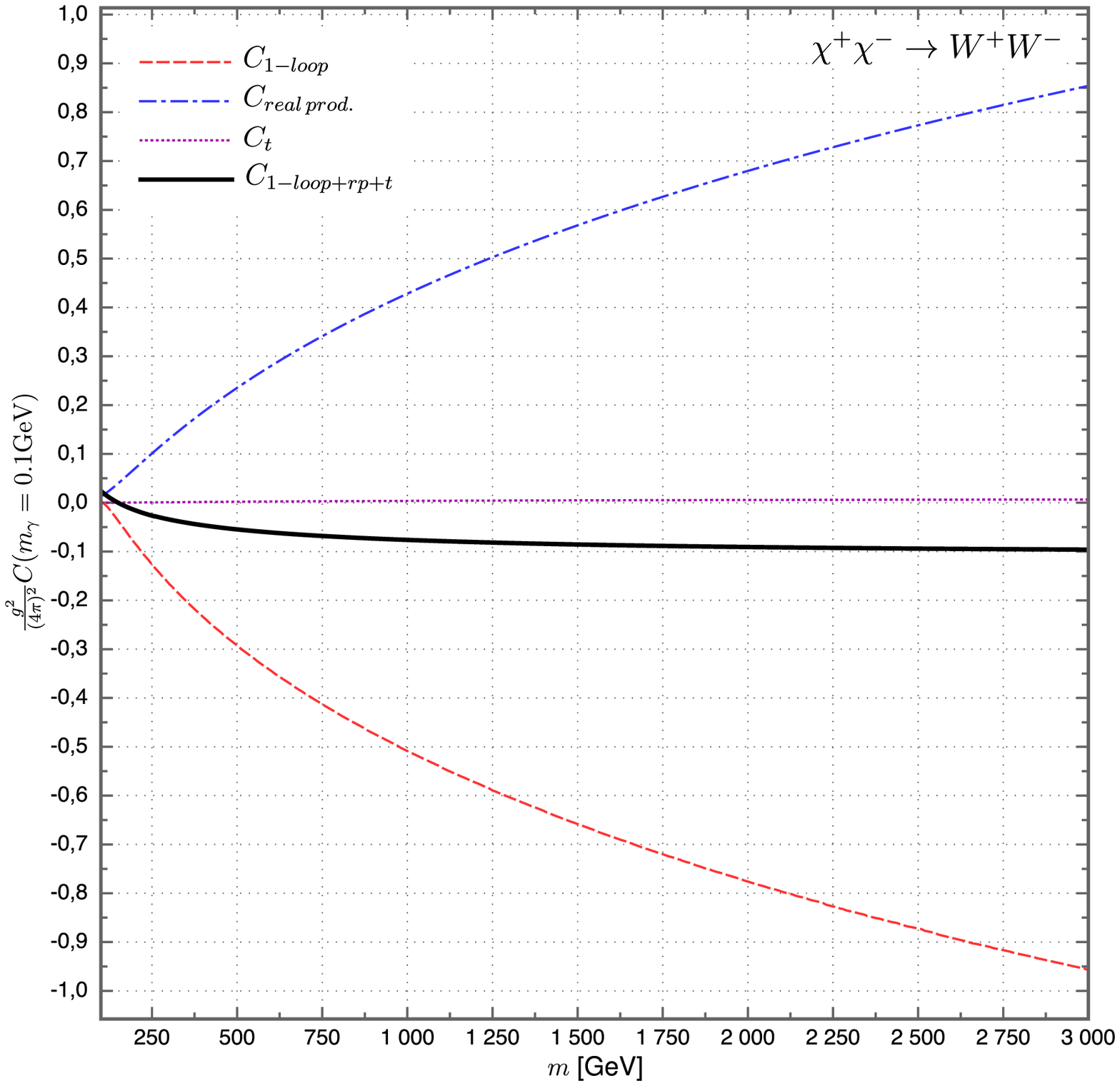} 
		\includegraphics[scale=0.4]{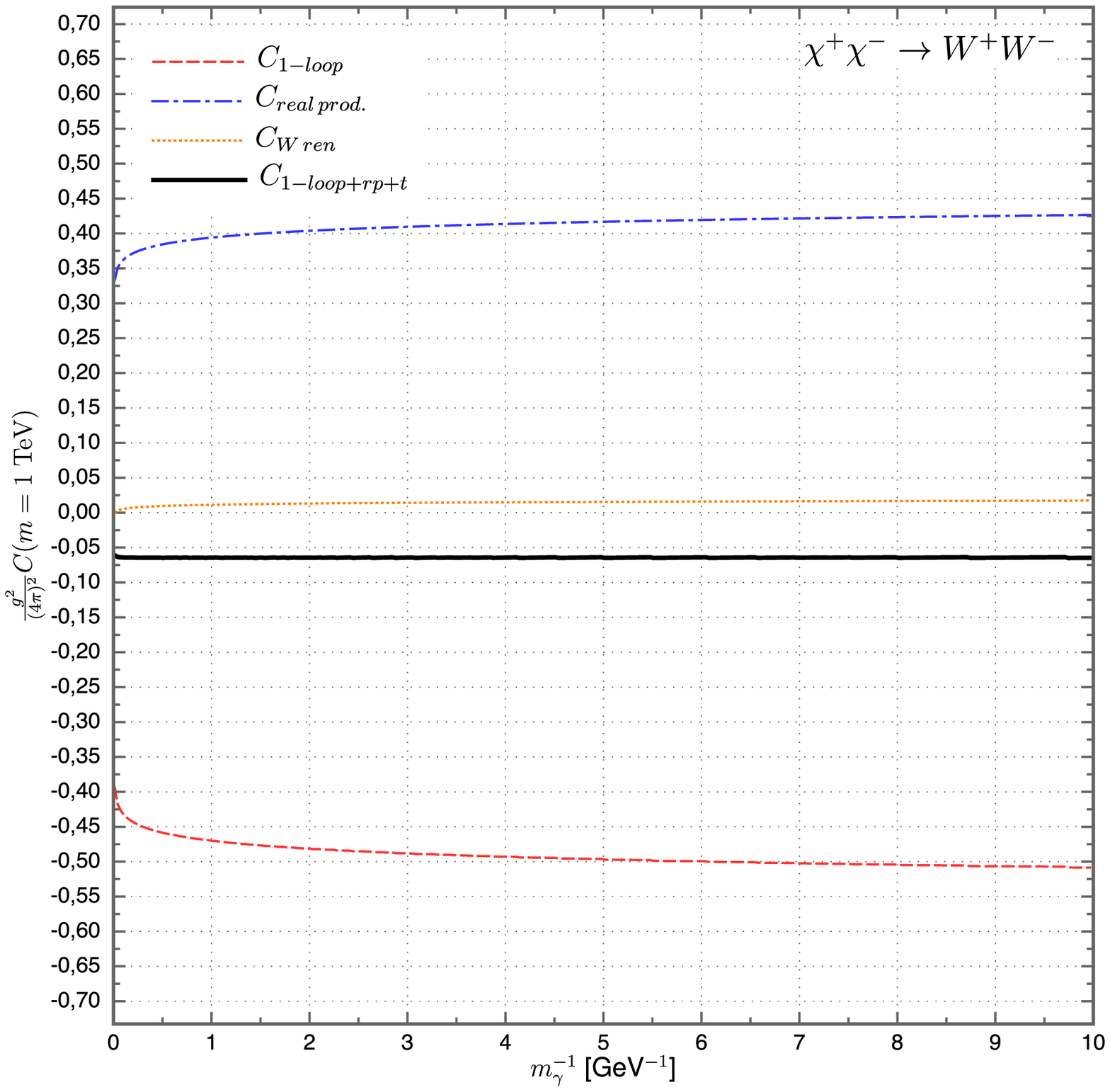} 
	\caption{The correction to the  $\chi^+\chi^-\to W^+W^-$ amplitude \textit{(left plot)} and dependence on the photon mass \textit{(right plot)}. The notation is the same as in figure \ref{plot_00}.}
    \label{plot_mc}
\end{figure}

On figure \ref{plot_mc} we show the full radiative correction to the amplitude of the process $\chi^+\chi^-\to W^+W^-$. When compared to the case of nuetralino annihilations, one immediately sees that although results are qualitatively similar, quantitatively are considerably smaller. In fact, the full one-loop result without including the one-loop Sommerfeld effect is within -10\% range even up to 3 TeV.

For completeness, also in this case we show the IR cancellation and that our results are independent of $m_\gamma$.

\subsection{The one-loop corrections to $\chi^+\chi^-\to ZZ,Z\gamma,\gamma\gamma$}

\begin{figure}[t]
\centering
	\includegraphics[scale=0.7]{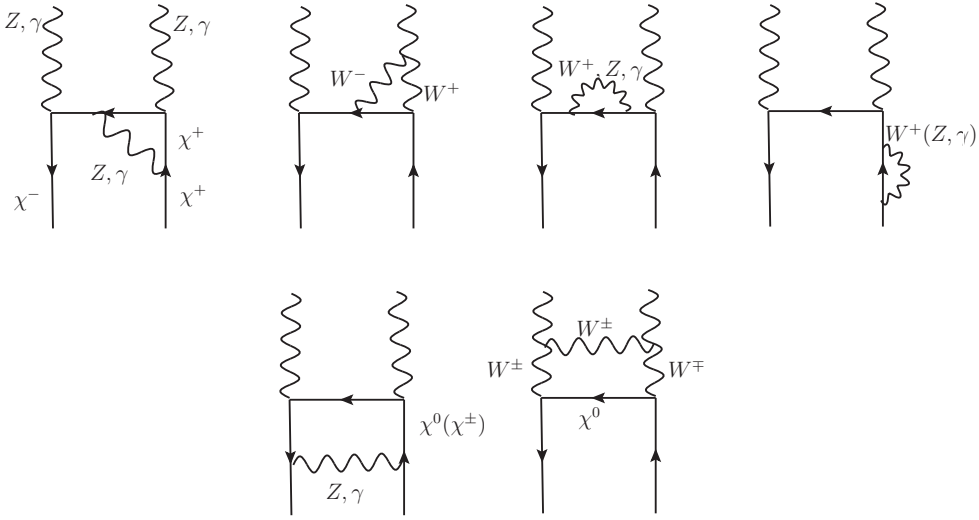} 
	\caption{The diagrams for the one-loop corrections to $\chi^+\chi^-$ annihilation to neutral gauge bosons.}
	\label{diagCW3}
\end{figure}

The diagrams to be computed are given on figure \ref{diagCW3}. In this case there is no wave-function renormalization of the final states, because they do not couple to the photon and thus do not exhibit IR divergences. Moreover, in this case there are no IR divergences in the total one-loop corrections, since the fermion wave-function renormalization cancels precisely the IR divergence coming from the correction to the initial states (the bottom left diagram).

There is also no three body production, since the emission of three $W^3$ (a mixture of $Z$ and $\gamma$) is forbidden by the CP conservation: the initial state being spin singlet has an even CP, while both $Z$ and $\gamma$ are CP-odd.\footnote{The processes involving two neutral gauge bosons one of them subsequently decaying into quark or lepton pairs are allowed, but similarly to what was said for the $t$ quark production they are very suppressed and therefore we neglect them.}
\begin{figure}[t]
\centering
	\includegraphics[scale=0.4]{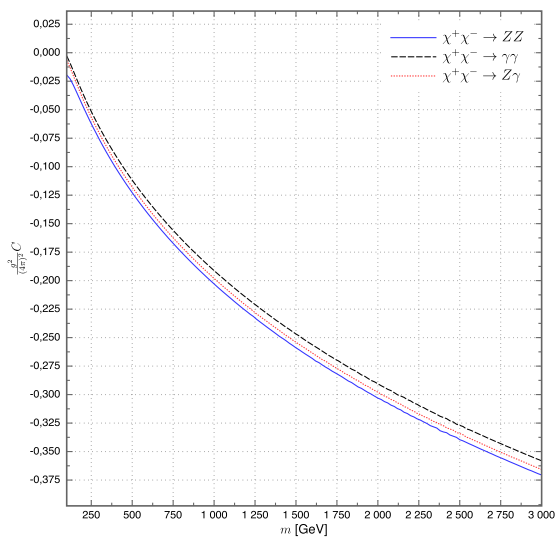}  
	\caption{The full one-loop corrections to the annihilation of $\chi^+\chi^-$ into $ZZ,Z\gamma,\gamma\gamma$. These corrections come only from the loop diagrams, because there is no real production in this case (see the text).}
    \label{plot_1loopW3W3}
\end{figure}

The results for the radiative correction to these processes are presented on figure \ref{plot_1loopW3W3}. The corrections are very similar to each other, as could be expected from the fact that since $m$ is much larger than $m_Z$, the differences in masses of the final states are not very important. On the other hand, the differences in couplings are taken into account in the tree level amplitudes for these processes (i.e. every of these three corrections is normalized to its own tree level amplitude).

 One can also see that the absolute value of these corrections is quite large, in fact considerably larger than for the annihilation into charged final states. This might look surprising, since there are less diagrams and none is IR divergent, but actually it can be easily understood by the fact that in this case there is no compensating effect of the real production.

\section{Final result for the Sommerfeld enhanced $\sigma v$}
\label{crossection}

Finally, we are ready to discuss the total annihilation cross sections of the $\chi^0\chi^0$ pair including both the radiative corrections discussed in the previous sections, as well as the full non-perturabative Sommerfeld enhancement. However, let us first make explicit which perturbation theory orders are included in our computation. The resulting cross-section $\sigma_2$ can be written as (see eqs. \eqref{sdef}, \eqref{eq:dsvgen} and \eqref{xsec2body}):
\begin{eqnarray}
\label{eq:order}
\sigma_2 v=\frac{1}{64\pi}\sqrt{1-\frac{m_g^2}{m^2}}\sum_{\rm pol} \biggl\lbrace &&|s_0|^2\left(|A_2^0|^2+2\, \Rea A_2^{0*} A_4^0 \right) + |s_\pm|^2\left(|A_2^\pm|^2+2 \,\Rea A_2^{\pm*} A_4^\pm\right)  \cr 
&&+ 2 \,\Rea\left(s_0^* s_\pm \left(  A_2^{0*} A_2^\pm  + A_4^{0*} A_2^\pm + A_2^{0*}  A_4^\pm \right) \right)\biggr\rbrace,
\end{eqnarray}
where $A^0_{(2,4)}$ and $A^{\pm}_{(2,4)}$ are the annihilation amplitudes,
  the superscript meaning initial $\chi^0\chi^0$ and $\chi^+\chi^-$ respectively, the subscript $(2,4)$ meaning the order $\mathcal{O}(g^2),\mathcal{O}(g^4)$,
i.e. $A_2$ is the $\mathcal{O}(g^2)$ and $A_4$ the $\mathcal{O}(g^4)$ part of $A$. The Sommerfeld factors $s_0$ and $s_\pm$ (see eq. \eqref{seampfact}) are evaluated non-perturbatively: the Sommerfeld enhanced terms  $s_0 A_2^0$ and $s_\pm A_2^\pm$  can be seen \cite{Iengo:2009ni} as the result of summing an infinite series of diagrams of increasing order in $g$, computed in the non-relativistic approximation.

In particular the perturbative series  for the Sommerfeld enhanced term $s_\pm A_2^\pm$ begins with diagram 4 in Fig. \ref{diag2} (in the non-relativistic approximation), which  together 
with diagram 5,  represented by  $A_4^\pm$ in eq. \eqref{eq:order}, makes possible the annihilation $\chi^0\chi^0\to ZZ,\gamma\gamma,Z\gamma$ to occur at  the  lowest order  $\mathcal{O}(g^4)$. Hence, formally the rate for this annihilation would be  $\mathcal{O}(g^8)$. However, the Sommerfeld factors are non-perturbatively evaluated and the 
resulting $s_0$ and $s_\pm$  are not perturbatively small, rather, in the TeV range, they are of order $1$ or much larger.

Our computation keeps in the cross-section all the terms up to $\mathcal{O}(g^6)$ included, besides the non-perturbative Sommerfeld factors $s_0$ and $s_\pm$.
Therefore, we include terms like $\Rea A_2^* A_4$ representing
the interference  of the diagrams $\mathcal{O}(g^4)$, which are not part of the Sommerfeld enhanced terms,  with diagrams $\mathcal{O}(g^2)$, but 
we do not include in $\sigma_2 v$ any terms of the order $\mathcal{O}(g^8)$, say $|A^0_4|^2$, or higher. They may play a role in the low mass region, especially for the annihilation to $ZZ$, $\gamma\gamma$ and $Z\gamma$, where the tree level is not present, if, in this mass region,  the Sommerfeld factor gives a result near to its perturbative expansion.
In our case at the TeV scale, they are subdominant. In order to incorporate them consistently with the Sommerfeld effect one would need to include also additional corrections, e.g. two-loop contribution to the $\chi^+\chi^-$ annihilation, which is beyond the scope of our paper.

\begin{figure}[t]
\centering
	\includegraphics[scale=0.4]{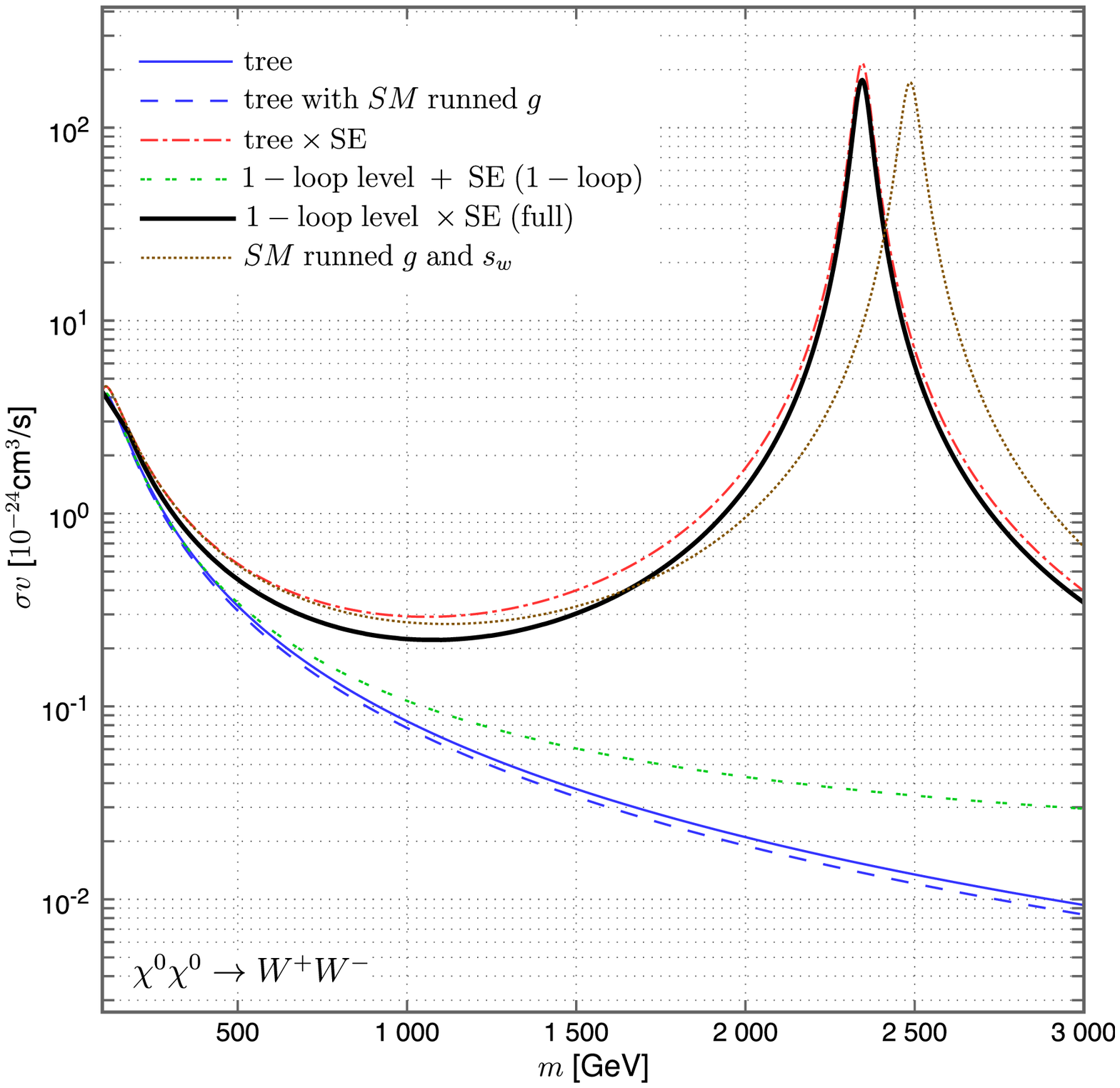}
		\includegraphics[scale=0.4]{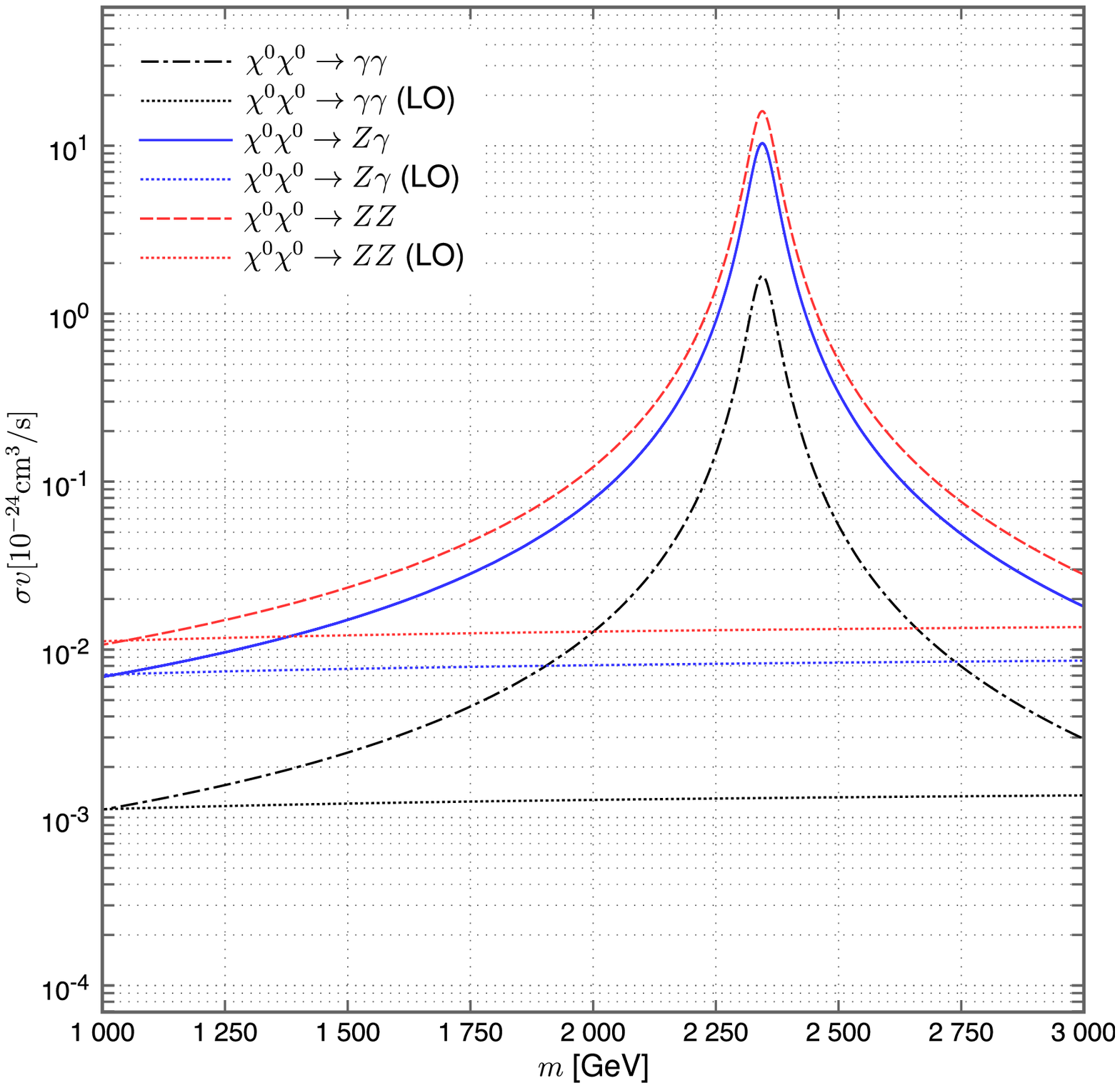}  
	\caption{\textit{Left plot}: the total cross-section for the annihilation of $\chi^0\chi^0$ to $W^+W^-$ (including the three body production). Our final results including both the one-loop corrections and the Sommerfeld effect (SE) are given by the solid black line. For comparison we plot the tree result (solid blue), tree level with the full SE (chain red), full one-loop level results but without non-perturbative SE (twin green) and the tree level with (dotted brown) and without SE (sparse blue) but with runned couplings at the scale $m$. \textit{Right plot}: the cross section for the annihilation to $ZZ,Z\gamma,\gamma\gamma$. The full one-loop results with the SE included are given and for comparison the leading order (LO) ones.}
    \label{plot_xsectfull}
\end{figure}

The results are presented on figure \ref{plot_xsectfull}. Let us first concentrate on the left plot showing the $\chi^0\chi^0\to W^+W^-$ annihilation process
(including the full Sommerfeld enhanced three body production cross-section, that is $\chi^0\chi^0\to W^+W^-\gamma$, $\chi^0\chi^0\to W^+W^-Z$ and $\chi^0\chi^0\to W^+W^-tb$, see section \ref{IRdiv}, which are $\mathcal{O}(g^6)$ and are added to $\sigma_2 v$).
Here we can extend our results below 1 TeV because the process $\chi^0\chi^0\to W^+W^-$ exists at tree level and thus neglecting terms of order $\mathcal{O}(g^8)$ does not introduce significant change. 
 Our full results including the one-loop corrections and the Sommerfeld effect are given by the solid black line. There is a clear resonance visible, which is due to the creation of a loosely bound state of the incoming neutralino pair. The resonance occurs approximately when the Bohr radius of the $\chi^0\chi^0$ pair matches the interaction range, i.e.:
\begin{equation}
\frac{1}{m\alpha}\approx \frac{1}{m_W}.
\end{equation}
For this reason, the position of the resonance depends strongly of the value of the coupling and this is why it is so important to use it at an appropriate scale. In section \ref{couplscale} we have discussed that the proper value is the one at the electroweak scale. If one uses instead its runned value at the scale $m$ (and do not include the radiative corrections) then one get the result plotted by the brown dotted line. That is, if one used the running of the couplings instead of doing full one-loop computation, one would get the resonance peak displaced in $m$ from about 2.35 to about 2.5 TeV. The shift of this peak influences also the mass of the Wino for which the correct thermal relic density is obtained (see ref. \cite{Hryczuk:2010zi}).

On the same plot we show that including only the one-loop approximation of the Sommerfeld effect is a good approximation only up to about 200 GeV, beyond which it breaks down, mainly due to the presence of the resonance. However, it is of course still more accurate than just using the tree level value. What might seem surprising is that using the tree level formula but with a running coupling constant at a scale $m$ (blue sparse line) is even a worse approximation than simply taking the standard tree level cross-section. This comes from the fact that running of the couplings captures only the UV effects of re-summation of large Log's (which gives a negative contribution), while in our setup the dominant correction to the annihilation amplitude is the Sommerfeld one, which is positive and (at a one-loop approximation) proportional to $m/m_W$.

We also do a comparison with the results using the full Sommerfeld corrections, but this time applied only to the tree level annihilation amplitudes (red chained line). The largest difference occurs before the resonance, for the masses of about 1 TeV, where the inclusion of one-loop contributions makes the full cross-section smaller by as much as about 30\%. In the resonance, although due to the logarithmic scale from the plot it seems that including the loop corrections do not change the result significantly, actually the radiative corrections make the value of $\sigma v$ in the peak is smaller by about 22\%.

On the right plot of figure \ref{plot_xsectfull} we present the results of the annihilation cross-section to $ZZ$, $Z\gamma$ and $\gamma\gamma$. Again a clear resonance is visible, for the same reason as before (in fact, by construction Sommerfeld effect is independent of the final states). However, the absolute value of the cross-section into $\gamma\gamma$
is about two orders of magnitude smaller than into $W^+W^-$, because the annihilation of $\chi^0\chi^0$ into neutral gauge bosons cannot occur at the tree level. In order to emphasize the importance of the full non-perturbative Sommerfeld effect, on the same plot we show the Leading Order (LO) results for those processes. They were computed retaining only the one-loop Sommerfeld correction, which makes the expected $m^{-2}$ dependence replaced by $m_W^{-2}$ one (see eq. \ref{eq:oneloopS}), leading to a result effectively independent of $m$.

In the Appendix \ref{append} we provide a fit to our results, listing in a Table \ref{tabfit} the numerical values for the parameters. This fit could be of use for possibly incorporating our results in further studies of the dark matter signals.

\section{Conclusions}
\label{conclusions}

In this paper we have computed the full annihilation cross-section including one-loop and the Sommerfeld electroweak corrections for the annihilation of the non-relativistic Majorana fermion being in the adjoint representation of the weak $SU(2)$. This scenario can be realized in number of physical situations, from which the most important is the neutralino in the MSSM in the limit in which it is purely Wino. The most interesting range of masses in this case is about a few TeV, because then the Wino can account for the thermal dark matter with the relic density in agreement with the observations. However, for such large values of the mass the perturbative computations are not sufficient. To give a correct annihilation cross-section in this case one needs to sum over all orders the ladder diagrams which give the Sommerfeld enhancement. 

In our work we included both these effects and discussed how to incorporate them simultaneously. In order to do so, one needs to compute both the $\chi^0\chi^0$ and $\chi^+\chi^-$ annihilation amplitudes at the same order, multiply them by corresponding Sommerfeld factors to get the corrected amplitude and finally take its modulus squared and integrate over the phase space. When computing the one-loop corrections one also has to be careful and do not include the one-loop version of the Sommerfeld correction, since it is included in the full non-perturbative treatment. 

The results we obtained introduce a relatively large one-loop corrections both to the $\chi^0\chi^0\to W^+W^-$ and $\chi^+\chi^-\to W^+W^-,ZZ,Z\gamma,\gamma\gamma$ processes. For a neutralino mass being $m=3$ TeV the corrections to $\sigma v$ reach more than -30\% (excluding the contributions of the Sommerfeld effect) for the neutralino annihilation,  -20\% for the $\chi^+\chi^-\to W^+W^-$ and more than -70\% for charginos annihilating into neutral gauge bosons. As we discussed in section \ref{chi0total}, this is due to the occurrence of powers of $\log m/m_W$, not related to UV divergences, that make the gauge theory to resemble a confining unbroken $SU(2)$.

On the other hand the Sommerfeld effect introduces a positive contribution, which already for masses of about 200 GeV overwhelms the other corrections and become far dominant when we enlarge the mass to the TeV scale. On top of that, the Sommerfeld factors exhibit a resonance due to forming a loosely bound state of the incoming particles. This has been already discussed in detail in literature (see e.g. \cite{Hisano:2002fk,Hisano:2003ec,Hisano:2004ds,Cirelli:2007xd}) and we obtain qualitatively the same results. However, since in this work we included rather large one-loop contributions, the precise values of the cross-section are slightly smaller, also in the resonance. Our final result for $\chi^0\chi^0\to W^+W^-$ cross-section before the resonance is up to about 30\% smaller than the one computed from the tree level cross-section multiplied by the Sommerfeld factor, whereas in the resonance it is smaller by about 22\%.

We also discuss the issue of the values of the couplings which we have to use, and argue that those should be the ones at the electroweak scale. This introduces a very mild modification to the one-loop corrections, but leads to a visible shift in the mass of the neutralino for which the resonance in the annihilation cross-section occurs (as well as in the relic density computations \cite{Hisano:2006nn,Cirelli:2007xd,Hryczuk:2010zi,Hryczuk:2011tq}). Our results show that the resonance takes place for $m\approx 2.35$ TeV whereas if we used the couplings runned up to the scale $m$ then this would shift to about 2.5~TeV.

The possible phenomenological applications of these computations are mainly concentrated on the modification of the cosmic ray spectra and therefore the predictions for the indirect detection. There inclusion of one-loop contributions introduce two effects. First, the overall cross-section being lowered by the one-loop corrections, both in the annihilations into $W^+W^-$ and $ZZ$, as well as $Z\gamma$ and $\gamma\gamma$. The two former processes contribute to the diffuse gamma ray spectra, while the two latter give a gamma line. Since our results show that the one-loop corrections are larger for the neutral final states than the charged ones, it suggests that the gamma lines are even fainter with comparison to the diffuse background. On the other hand, the second consequence of the full one-loop computation is that now the total cross-section incorporates also the real production, which may alter the final spectra significantly (see e.g. \cite{Kachelriess:2009zy,Ciafaloni:2010qr,Bell:2011eu,Ciafaloni:2011sa}). We leave the full discussion of this point to a future work \cite{inprep}. 

\acknowledgments
We would like to thank Piero Ullio for useful discussions and encouragements and also Bobby Acharya for pointing out a normalization mistake.

\appendix
\section{Fit to the full Sommerfeld enhanced $\sigma v$}
\label{append}

In order to make our results easy to use in further computations, we also provide a fit of the full cross-sections (comprising the three body production) including the one-loop corrections and the full Sommerfeld effect. For the masses above 1 TeV all of them are fitted with the same functional form:
\begin{equation}
f_>(m)=\frac{b_0+b_1 m + b_2 m^2 + b_3 m^3}{c_0+c_1 m + c_2 m^2 + c_3 m^3},
\end{equation}
while below 1 TeV for the cross-section to $W^+W^-$:
\begin{equation}
f_<(m)=\frac{1}{m^2}\sum_{n=0}^4 a_n \log^n\left(\frac{m}{m_W}\right).
\end{equation}
Therefore, 
\begin{equation}
\sigma v_{\rm fit}^{WW}=\biggl\lbrace
\begin{array}{ll}
f_<(m)	& \quad m\leq 1000\, {\rm GeV} \\
f_>(m)	& \quad m>1000\, {\rm GeV}
\end{array}
,\qquad\sigma v_{\rm fit}^{ZZ,Z\gamma,\gamma\gamma}|_{m>1\;{\rm TeV}}=f_>(m).
\end{equation}

\begin{figure}[t]
\centering
	\includegraphics[scale=0.7]{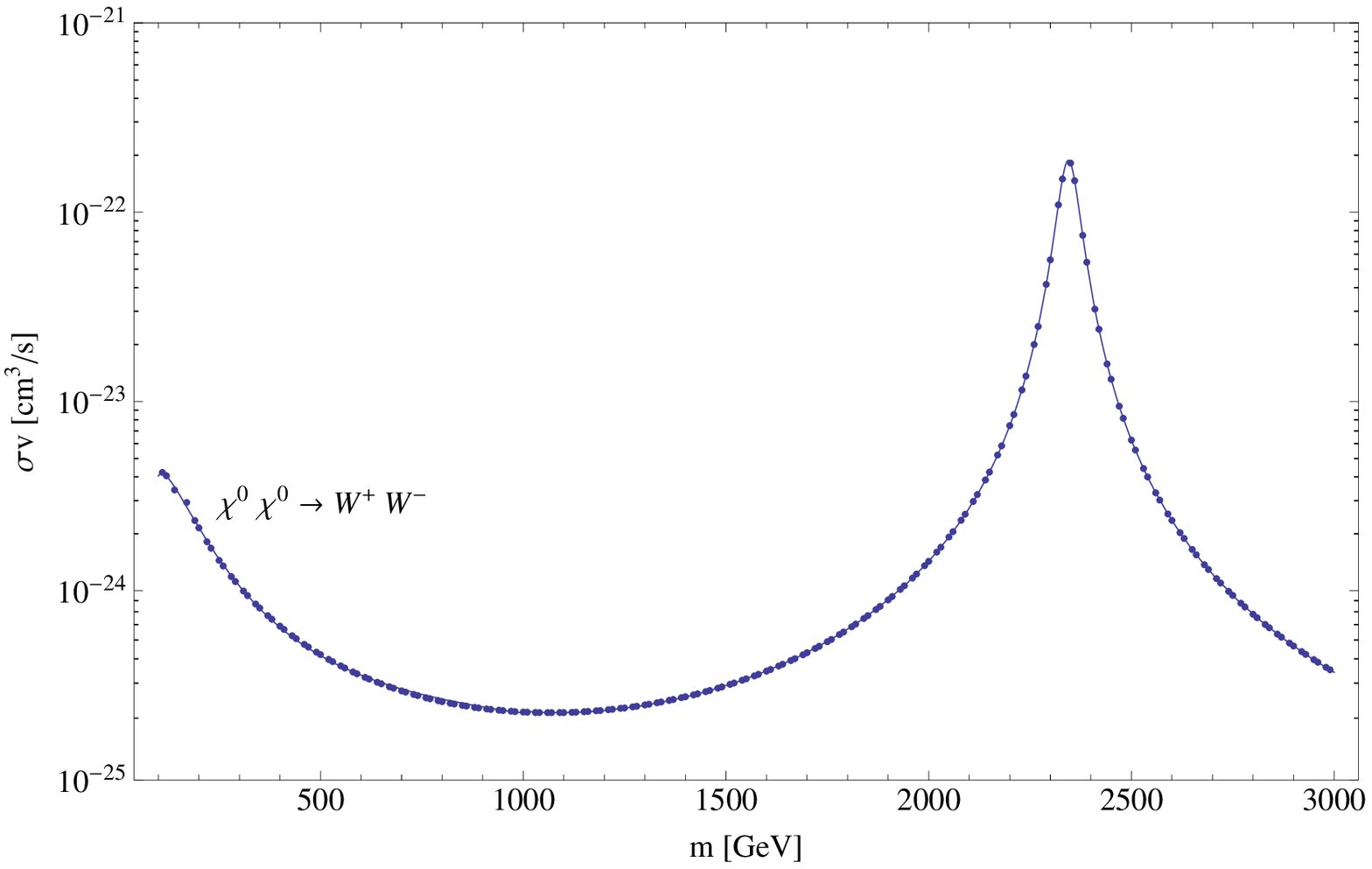} 
	\includegraphics[scale=0.65]{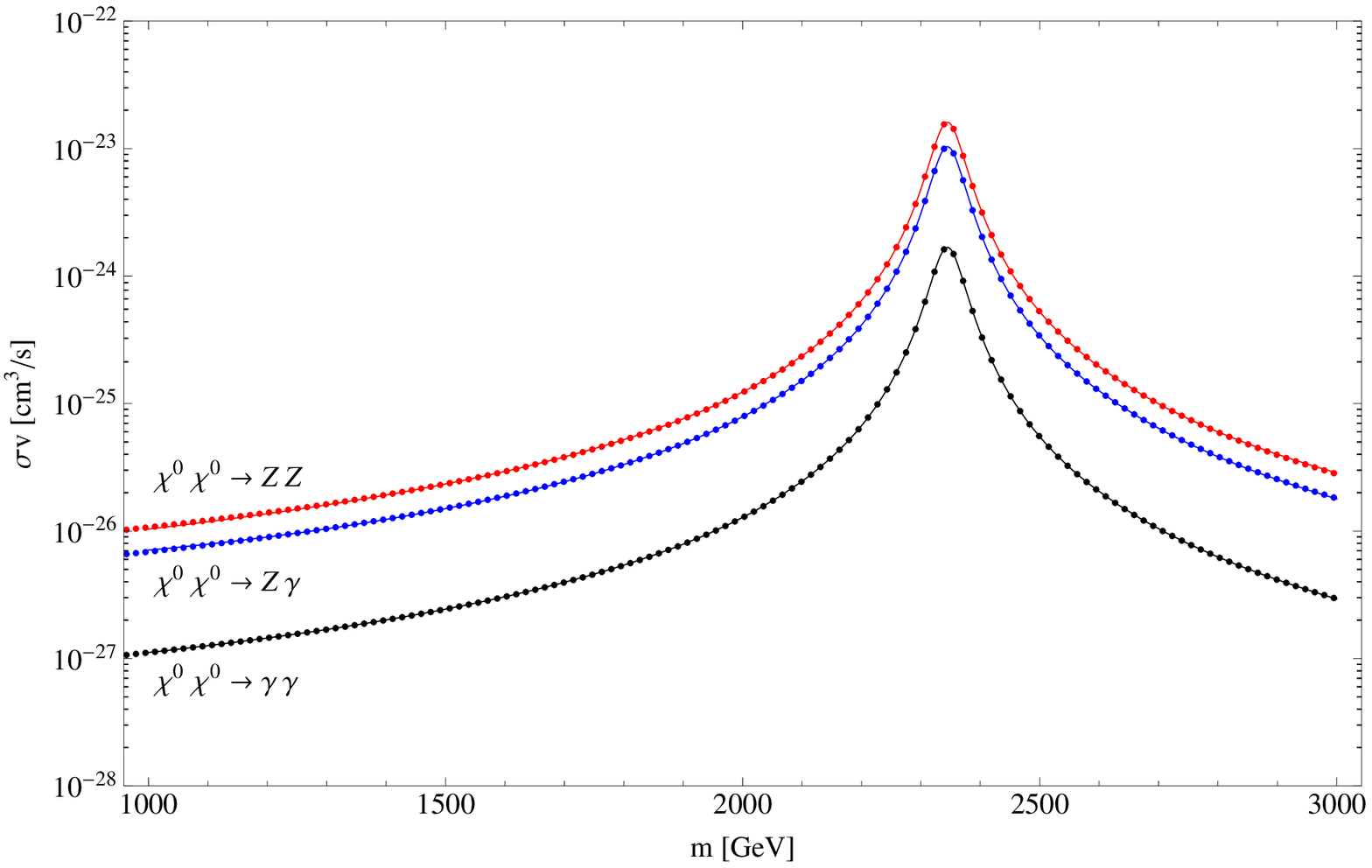} 
	\caption{The comparison between the numerical results (points) and the fit (solid lines) for the annihilation cross-section of $\chi^0\chi^0$ to $W^+W^-$ \textit{(top)} and neutral gauge bosons \textit{(bottom)}. The accuracy of the fits is within 1\%.}
	\label{plot_fit}
\end{figure}

The values for the coefficients providing the best fit are given in Table \ref{tabfit}, for the various annihilation channels. Both $f_{<,>}(m)$, with $m$ and $m_W$ expressed in GeV, give the result in units of $\rm cm^3/s$. The difference between these fits and the numerical results are within 1\% range (see figure~\ref{plot_fit}).

{\linespread{1.1}
\begin{table}[tb]
\begin{tabular}{|c|c|c|c|c|}
\hline \multicolumn{5}{|c|}{$\chi^0\chi^0\to W^+W^-$} \\
\hline  $\bf a_0$ & $\bf a_1$ & $\bf a_2$ & $\bf a_3$ & $\bf a_4$ \\
\hline $1.32188\times 10^{-20}$ & $1.36268\times 10^{-19}$ & $-6.24676\times 10^{-20}$ & $-8.02201\times 10^{-21}$ & $9.73209\times 10^{-21}$\\ 
\hline $\bf b_0$ & $\bf b_1$ & $\bf b_2$ & $\bf b_3$ & \\ 
\hline $2.67711\times 10^{-12}$ & $1.35313\times 10^{-15}$ & $-7.00791\times 10^{-19}$ & $2.49388\times 10^{-22}$ &  \\
\hline $\bf c_0$ & $\bf c_1$ & $\bf c_2$ & $\bf c_3$ & \\ 
\hline $-5.26864\times 10^{13}$ & $1.44622\times 10^{11}$ & $-9.46014\times 10^7$ & $1.81301\times 10^{4}$ &  \\ 
\hline
\bottomrule
\hline \multicolumn{5}{|c|}{$\chi^0\chi^0\to ZZ$} \\
\hline $\bf b_0$ & $\bf b_1$ & $\bf b_2$ & $\bf b_3$ & \\ 
\hline $3.3929\times 10^{-15}$ & $-7.98612\times 10^{-19}$ & $-1.2351\times 10^{-22}$ & $4.67818\times 10^{-26}$ &  \\
\hline $\bf c_0$ & $\bf c_1$ & $\bf c_2$ & $\bf c_3$ & \\ 
\hline $8.52321\times 10^{11}$ & $-8.39761\times 10^8$ & $2.51261\times 10^5$ & $-2.05257\times 10^1$ &  \\ 
\hline
\bottomrule
\hline \multicolumn{5}{|c|}{$\chi^0\chi^0\to Z\gamma$} \\
\hline $\bf b_0$ & $\bf b_1$ & $\bf b_2$ & $\bf b_3$ & \\ 
\hline $-1.80836\times 10^{-15}$ & $3.37165\times 10^{-18}$ & $-7.40609\times 10^{-22}$ & $7.48779\times 10^{-26}$ &  \\
\hline $\bf c_0$ & $\bf c_1$ & $\bf c_2$ & $\bf c_3$ & \\ 
\hline $-7.32662\times 10^{11}$ & $1.74234\times 10^{9}$ & $-1.0862\times 10^6$ & $2.03233\times 10^2$ &  \\ 
\hline
\bottomrule
\hline \multicolumn{5}{|c|}{$\chi^0\chi^0\to \gamma\gamma$} \\
\hline $\bf b_0$ & $\bf b_1$ & $\bf b_2$ & $\bf b_3$ & \\ 
\hline $4.00871\times 10^{-12}$ & $-2.48789\times 10^{-15}$ & $5.60019\times 10^{-19}$ & $-5.44691\times 10^{-23}$ &  \\
\hline $\bf c_0$ & $\bf c_1$ & $\bf c_2$ & $\bf c_3$ & \\ 
\hline $8.04594\times 10^{15}$ & $-9.37744\times 10^{12}$ & $3.60859\times 10^9$ & $-4.57474\times 10^5$ &  \\ 
\hline
\end{tabular} 
\caption{The obtained values of the parameters giving the best fit for the cross-sections $\sigma v_{\rm fit}$ for all $\chi^0\chi^0$ annihilation channels.}
\label{tabfit}
\end{table} }

\clearpage

\bibliographystyle{amsplain}

\end{document}